\documentclass[12pt,preprint]{aastex}

\shorttitle{Absorption from Interstellar C$_2$}
\shortauthors{Hupe et al.}

\begin{document}

\title{Ultraviolet Measurements of Interstellar C$_2$}

\author{Ryan C. Hupe\altaffilmark{1}$^,$\altaffilmark{2},
Y. Sheffer\altaffilmark{1}$^,$\altaffilmark{3}, and 
S.R. Federman\altaffilmark{1}}

\altaffiltext{1}{Department of Physics and Astronomy, University of Toledo,
Toledo, OH 43606, USA; steven.federman@utoledo.edu}
\altaffiltext{2}{Department of Physics, Ohio State University, Columbus, 
OH 43210, USA; hupe@mps.ohio-state.edu}
\altaffiltext{3}{Department of Astronomy, University of Maryland, 
College Park, MD 20742, USA; ysheffer@astro.umd.edu}

\begin{abstract}
We analyzed archival spectra acquired with the {\it Hubble Space Telescope} 
for a study of interstellar C$_2$.  Absorption from the electronic 
transitions, $D$ $^1\Sigma^+_{\rm u}$ -- $X$ $^1\Sigma^+_{\rm g}$ (0,0) as 
well as $F$ $^1\Pi_{\rm u}$ -- $X$ $^1\Sigma^+_{\rm g}$ (0,0) and (1,0), 
was the focus of the study.  Our profile syntheses revealed that the lines 
of the $F-X$ bands were broadened as a result of a perturbation involving 
the upper levels.  Further evidence for the perturbation came from 
anomalies in line strength and position for the $F-X$ (0,0) band.  The 
perturbation likely arises from a combination of triplet-singlet interactions 
involving spin-orbit mixing between $^3\Pi_{\rm u}$ states and 
$F$ $^1\Pi_{\rm u}$ and an avoided crossing between the $^3\Pi_{\rm u}$ 
states.  Tunneling through a potential barrier caused by the 3 and 
4 $^1\Pi_{\rm u}$ states and spin-orbit mixing with other close-lying 
triplet states of $ungerade$ symmetry are less likely.  
Except for the broadening, lines in the $F-X$ (1,0) band appear free from 
anomalies and can be used to study interstellar C$_2$; new results for 10 
sight lines are presented.
\end{abstract}


\section{Introduction}

The C$_2$ molecule is a useful diagnostic of the physical conditions in 
diffuse molecular clouds.  Analysis of its rotational excitation provides 
information on kinetic temperature, density of collision partners 
[$n_{coll}$ $=$ $n$(H~{\small I}) $+$ $n$(H$_2$)], and the flux of near 
infrared radiation permeating the gas (van Dishoeck \& Black 1982).  Because 
it acts as an intermediary in the chemical sequence leading to CN in these 
clouds (Federman et al. 1984; Federman et al. 1994), modeling 
this chemistry allows one to infer the total gas density [$n_H$ $=$ 
$n$(H~{\small I})~$+$~2~$n$(H$_2$)] and the ultraviolet (UV) flux incident 
on the cloud.  Many studies have extracted this information from 
measurements of absorption from the series of bands in the $A-X$ Phillips 
system (e.g., Chaffee et al. 1980; Hobbs \& Campbell 1982; Danks \& 
Lambert 1983; van Dishoeck \& de Zeeuw 1984; van Dishoeck \& Black 1989; 
Federman et al. 1994; Sonnentrucker et al. 2007; Ka\'{z}mierczak et al. 
2010).

The molecule also reveals absorption at near and far UV wavelengths, 
which is the focus of our work.  Using the $Copernicus$ satellite, Snow 
(1978) reported a tentative detection of absorption from the R(0) line 
of the Mulliken $D-X$ (0,0) band near 2313 \AA\ toward $\zeta$ Oph.  
Observations with the {\it International Ultraviolet Explorer} revealed 
absorption from the $F-X$ (0,0) band at 1342 \AA\ (Lien 1984) 
toward X Per, but the $D-X$ (0,0) and $F-X$ (1,0) bands were not seen.  
Lien (1984) suggested that the oscillator strength ($f$-value) for the 
$F-X$ (0,0) band was at least 0.10.  Lambert et al. (1995) 
detected absorption from the two (0,0) bands toward $\zeta$ Oph, using the 
Goddard High Resolution Spectrograph (GHRS) on the 
{\it Hubble Space Telescope} ($HST$).  
They inferred an $f$-value for the $F-X$ (0,0) band ($f_{00}^{F-X}$) of 
$0.10\pm0.01$ based on a comparison of the amount of absorption relative 
to that seen from the $D-X$ (0,0) band.  A theoretical value of 0.0545 
for $f_{00}^{D-X}$ (Chabalowski et al. 1983; Bruna \& 
Wright 1992) was adopted for the comparison.  A surprising finding from 
their analysis was that Lambert et al. had problems fitting the band 
profile.  The P(6) and Q(12) lines were too weak, the P(8) and Q(14) lines 
were absent, and the amount of absorption was too strong between 1342.5 
and 1343.0 \AA.  We note that absorption from only even levels is 
observed.

Subsequent analysis of $HST$ observations revealed further information 
about the UV transitions in C$_2$.  Kaczmarczyk (2000) examined absorption 
toward X Per from the $F-X$ (0,0) band and found additional line anomalies.  
Starting from the results from the $D-X$ band, Kacmarczyk showed that 
R($J^{\prime\prime}>2$), Q(8), Q(10), and P(4) were too 
weak.  In a detailed study, 
Sonnentrucker et al. (2007) expanded upon this work.  First, they 
obtained a better fit to line positions for the $D-X$ band by adopting 
$\nu_0$ $=$ 43227.2 cm$^{-1}$ from Sorkhabi et al. (1998).  They also 
noted that a shift in line position occurred for 
lines involving $J^{\prime\prime}$ $<$ 
10 for the $F-X$ (0,0) band.  Sonnentrucker et al. pointed out that the 
anomalies seen in the $F-X$ (0,0) band did not arise from stellar 
features in the background continuum because the stars in their sample had 
different spectral types, $v$ sin$i$, etc.  In the end, the fits to the 
$D-X$ (0,0) and $F-X$ (1,0) bands were very good, when $f_{10}^{F-X}$ of 
about 0.06 was used. 

Through further analysis, we refine our knowledge of the perturbation(s) 
responsible for the line anomalies.  As in Lambert et al. (1995) and 
Sonnentrucker et al. (2007), we start with syntheses of the $D-X$ band.  
Since we also find fits to the $F-X$ (1,0) band that are quite good, we 
extract C$_2$ column densities for directions where we obtained CO 
abundances from the far UV spectra previously (Pan et al. 2005; Sheffer 
et al. 2008).  The data for our study are described in \S~2, as is our 
method of processing the spectra, and \S~3 provides the results of our 
analyses.  They are discussed in \S~4.  We summarize our findings in the 
final Section.

\section{Data and Their Analysis}
 
The spectra were downloaded from the $HST$ archive 
at the Multiwavelength Archive at the Space Telescope Science Institute 
(MAST) as part of our earlier studies.  In particular, we used the 
GHRS spectra described by Lambert et al. (1995), but analyzed it together 
with spectra from the Space Telescope Imaging Spectrograph (STIS) in a 
self-consistent manner.  Far UV spectra acquired with STIS for our CO 
studies (Pan et al. 2005; Sheffer et al. 2008) comprised the bulk of the 
sample.  As for the near UV spectra obtained 
with STIS that contain absorption from the $D-X$ band, 
we downloaded the same exposures as Sonnentrucker et al. (2007), but 
analyzed them in an independent manner.  Our analyses are based on 
rectified spectra, where the stellar continuum was removed through fits 
involving low-order polynomials in most cases.  In some cases, such as for 
the $F-X$ bands seen in the spectrum of HD~203532, stellar features had 
to be fitted and removed as well.

The initial focus was on the directions with absorption from all three 
bands -- HD~24534 (X Per), HD~27778 (62 Tau), HD~147888 ($\rho$ Oph D), 
HD~149757 ($\zeta$ Oph), and HD~207198.  The spectra of $\zeta$ Oph were 
from GHRS observations.  Because the spectra for X Per revealed strong 
absorption and had high signal to noise, they were analyzed first; 
analysis of the spectra for the four other sight lines provided 
confirmation of our results regarding the perturbations involving the 
$F-X$ transitions.  In light of the ability of Lambert et al. (1995) and 
Sonnentrucker et al. (2007) to fit the $D-X$ band with confidence, we 
synthesized this band first, basing it on resolving powers of 80,000 for 
the GHRS spectrum and about 100,000 for STIS and on an $f$-value of 
0.0545.  The component structures used in the syntheses come 
from our earlier work (Lambert et al. 1995; Sheffer et al. 2008).  
Pan et al. (2005) described the close 
correspondence in component structure for CO and CN, providing evidence 
that these two molecules coexist in the core of diffuse molecular 
clouds.  Chemical studies (e.g., Federman et al. 1994) and analyses of 
CO and C$_2$ excitation (Sheffer et al. 2007) indicate that the 
C$_2$ molecule is also in the core.  For comparison, Sonnentrucker et al. 
(2007) synthesized the C$_2$ transitions through a combination of the 
component structure seen in K~{\small I}, CH, and CN absorption.

We adopted the line lists given by Sonnentrucker et al. (2007) for our 
profile syntheses with the code ISMOD (see \S2.5 in Sheffer et al. 2008).  
We also found that improved fits to the $D-X$ band arise when the constants 
from Sorkhabi et al. (1998) are adopted.  The fits to the $D-X$ band 
yielded column densities for individual rotational levels of the molecule.  
These column densities then were applied to the $F-X$ bands, where we 
allowed the band $f$-value and line width to vary.  By allowing the 
line width to vary, we were able to search for intrinsic broadening 
associated with the perturbations seen in earlier studies.  An example 
of our fits to the $D-X$ and $F-X$ bands toward 62 Tau (HD~27778) is shown 
in Figures 1 and 2, respectively.  Once a self-consistent set of 
$f$-values and intrinsic line broadening for the $F-X$ bands were found, 
we were able to extract C$_2$ column densities for additional sight lines.  
Examples appear in Fig. 3.

\section{Results}

\subsection{$F-X$ Bands}

We confirm earlier astronomical analyses of these bands.  We find that the 
$f$-values for the (0,0) band (0.10, Lambert et al. 1995) and (1,0) band 
(0.06, Sonnentrucker et al. 2007) provide a set of self-consistent oscillator 
strengths when used in combination with the $D-X$ (0,0) and $A-X$ (2,0) 
bands.  We also observe the line anomalies noted by Lambert et al. (1995), 
Kacmarczyk (2000), and Sonnentrucker et al. (2007).

We go beyond this body of work in two important ways, based mainly on 
absorption toward X Per whose spectra have the highest signal to noise.  
First, we are best able to fit the lines in the two $F-X$ bands with a 
resolving power of 60,000, significantly lower that the 
resolving power achievable with STIS.  
We believe this results from increased intrinsic broadening 
arising from shortened lifetimes in levels of the $F$ state.  A lifetime 
of about $6 \times 10^{-12}$ s$^{-1}$, compared to typical radiative 
lifetimes of ns (see below), is inferred.  Second, while Sonnentrucker 
et al. (2007) found that the $J^{\prime\prime}$ $<$ 10 lines of the 
$F-X$ (0,0) band are shifted by about 36 m\AA\ (or 8 km s$^{-1}$), 
we further quantify details about the shifts.  We focus on the upper 
rotational levels where the perturbation is taking place.  We find 
the following offsets in velocity (km s$^{-1}$) as a function of 
$J^{\prime}$.

\noindent
\resizebox{6.5in}{!}{
\begin{tabular}{lcccccccccccccccccc}
$J^{\prime}$ & 1 & 2 & 3 & 4 & 5 & 6 & 7 & 8 & 9 & 10 & 11 & 12 & 13 & 
14 & 15 & 16 & 17 & 18 \\
offset & $+8.0$ & $+8.0$ & $+8.0$ & $+8.0$ & $+8.5$ & $+9.0$ & $+9.5$ & 
$+8.0$ & $+11.5$ & $+15.0$ & $+15.0$ & $-2.0$ & $0.0$ & $+2.0$ & 
$\ldots$ & $0.0$ & $\ldots$ & $0.0$ \\
(km s$^{-1}$) & \\
\end{tabular}
}

\noindent
All levels with $J^{\prime}$ $\le$ 11 show substantial shifts, with 
$J^{\prime}$ of 10 and 11 shifted by 5 resolution elements.  No shift is 
seen for J$^{\prime}$ $\ge$ 12.

\subsection{Interstellar C$_2$}

The results of our syntheses appear in Tables 1 and 2.  Table 1 gives 
column densities for individual rotational levels for the directions 
with $D-X$ spectra and comparisons to earlier analyses (Lambert et al. 
1995; Kaczmarczyk 2000; Sonnentrucker et al. 2007), while Table 2 
provides column densities for new sight lines.  As expected, essentially 
all our results in Table 1 agree with the earlier determinations at the 
1-$\sigma$ level.  The largest differences appear for the $J$ $=$ 2 and 4 
levels toward X Per, where our values are about 2-$\sigma$ larger.  We 
also report tentative detections for $J$ $=$ 16 and 18 toward X Per and 
$\rho$ Oph D.

\section{Discussion}

\subsection{$D-X$ Band}

Before discussing the perturbations in the $F-X$ bands, we summarize the 
situation for the Mulliken band.  The first studies of the radiative 
properties of C$_2$ molecules in the $D-X$ (0,0) band yielded laboratory 
lifetimes for the upper level.  Using the phase shift method, Smith 
(1969) measured a lifetime of $14.6\pm1.5$ ns for an $f$-value of 
$0.055\pm0.06$.  Curtis et al. (1976) obtained a lifetime of $18.1\pm1.0$ 
ns on an unresolved group of lines with the High Frequency Deflection 
Technique.  Subsequent studies relied on large-scale theoretical 
calculations.  Chabalowski et al. (1983) performed a multireference 
double excitation -- configuration interaction (MRD-CI) calculation that 
yielded a lifetime of 14 ns and an $f$-value of 0.054.  The computation 
by Bruna \& Wright (1992), also based on MRD-CI, gave a lifetime of 14.6 
ns and $f$-value of 0.0546.  Kokkin et al. (2007) used multireference 
configuration interaction (MRCI) techniques and found 
a ratio of oscillator strengths, $f_{20}^{A-X}$/$f_{00}^{D-X}$, 
of 0.0212 versus the ratio quoted by Lambert 
et al. (1995) of $0.0226\pm0.029$, which was based on $f_{00}^{D-X}$ $=$ 
0.0545.  Most recently, Schmidt \& Bacskery (2007) enlarged the scope of 
calculations performed by Kokkin et al. (2007) by incorporating 
non-orthogonal orbitals.  Schmidt \& Bacskery obtained a lifetime of 
13.4 ns, an $f$-value of 0.0535, and a ratio of 0.0266.  The value for 
$f_{00}^{D-X}$ seems quite secure.

\subsection{$F-X$ Bands}

The only laboratory-based spectroscopic study is that of Herzberg et al. 
(1969), whose molecular constants are used for analysis of interstellar 
absorption.  The derivation of these constants relied on lines with 
$J^{\prime\prime}$ $\ge$ 10 for the most part.  For the (0,0) band, 
laboratory data do not exist for lower values of $J^{\prime\prime}$, 
while data only on the Q branch of the (1,0) band are presented by 
Herzberg et al. (1969) for these rotational levels.  
Therefore, the only quantitative information 
for the perturbations involving the $F-X$ levels at the present time 
is based on astronomical spectra -- broadened lines and line shifts of 
8 to 15 km s$^{-1}$.

Several quantum mechanical calculations examined Rydberg states, such 
as $F$ $^1\Pi_{\rm u}$.  The first one (Barsuhn 1972), which was based 
on self-consistent field and configuration interaction (SCF$+$CI), 
investigated the energies for the Rydberg states.  Another SCF$+$CI 
calculation by Pouilly et al. (1983) focused on the C$_2$ photodissociation 
rate for chemical models of interstellar clouds and comets.  They also 
determined a value for $f_{00}^{F-X}$of 0.02, which differs from 
$0.10\pm0.01$ derived by Lambert et al. (1995) from interstellar spectra.  
The most comprehensive theoretical study to date (Bruna \& Grein 2001) 
involved MRCI calculations of valence and Rydberg states between 7 and 
10 eV.  Bruna \& Grein obtained $f_{00}^{F-X}$ of 0.098, in much better 
agreement with astronomical determinations.  We are not aware of any 
reported calculation giving a value for $f_{10}^{F-X}$.

Our results suggest that the $v$ $=$ 0 level of the $F$ $^1\Pi_{\rm u}$ 
state is affected more than the $v$ $=$ 1 level because both line 
anomalies (in strength and position) and 
broadening are seen only in lines involving $v$ $=$ 0.  We examine the 
theoretical results of Bruna \& Grein (2001) in order to interpret 
the interstellar observations.  They provide the following 
description of the $F$ (2 $^1\Pi_{\rm u}$) state.  Its potential energy 
curve is affected by avoided crossings at about 2.8 Bohr with the curves 
associated with 3 $^1\Pi_{\rm u}$ and 4 $^1\Pi_{\rm u}$, yielding 3 bound 
vibrational levels for the $F$ state.  We note that Herzberg et al. 
(1969) observed the lower two levels, which are also the only ones 
associated with interstellar spectra.  The depth of the potential well is 
about 0.5 eV.  The $F$ state not only is metastable 
regarding radiative emission, but it is also metastable with respect to 
predissociation.  The latter arises because the minimum in the potential 
curve for the $F$ state lies about 1.50 eV above its diabatic limit, 
C[$^1$D] $+$ C[$^1$D].  The $F$ state is somewhat unique since the 
same potential curve undergoes two very different predissociation 
mechanisms before and after the minimum.

The likely cause for the perturbations seen by Lambert et al. (1995), 
Kaczmarczyk (2000), Sonnentrucker et al. (2007), and us involves 
triplet-singlet interactions, via spin-orbit mixing, and an avoided crossing.  
According to Bruna \& Grein (2001), the curves for 2 and 3 $^3\Pi_{\rm u}$ 
are essentially repulsive.  The inner portion of the 3 $^3\Pi_{\rm u}$ 
curve constitutes the triplet counterpart of $F$ $^1\Pi_{\rm u}$.  
There exists an avoided crossing between the curves for 
2 and 3 $^3\Pi_{\rm u}$ at approximately 9.2 eV near 
2.4 Bohr that may affect the $F$ $^1\Pi_{\rm u}$ potential.  Beyond the 
avoided crossing, the curve for 3 $^3\Pi_{\rm u}$ lies close to that 
for $F$ $^1\Pi_{\rm u}$ for a short interval.  

Portions of the potential curves for 4 and 5 $^3\Pi_{\rm u}$ are also 
located nearby (around 9.5 eV and internuclear separation of about 
2.6 Bohr).  Both curves have potential 
wells; Bruna \& Grein (2001), however, do not list possible vibrational 
levels for 4 and 5 $^3\Pi_{\rm u}$. If they existed, triplet-singlet 
mixing may arise, but the line broadening seen by us suggests 
predissociation is taking place and that would require another avoided 
crossing with a repulsive curve.

Tunneling through the potential barrier created by avoided crossings among 
$^1\Pi_{\rm u}$ states and spin-orbit mixing between $F$ $^1\Pi_{\rm u}$ 
and other close-lying triplet states of $ungerade$ symmetry could also 
lead to the perturbations seen in interstellar spectra (P. Bruna, private 
communication), but are less likely.  Tunneling is expected to affect 
the $v$ $=$ 2 level the most because it lies near the top of the barrier; 
such an effect may explain why lines from this level are not seen in 
laboratory (Herzberg et al. 1969) or interstellar spectra.  The spin-orbit 
mixing considered here involves repulsive states that can predissociate the 
$F$ state.  We consider the results of CI calculations by Kirby \& Liu (1979).  
Their Table VI lists weakly bound states that may be of relevance here.  
Two states, 2 $^3\Sigma^+_{\rm u}$ and 1 $^3\Sigma^-_{\rm u}$, 
may perturb the $F$ state in a significant way.  However, in a theoretical 
analysis of the interaction between the $X$ $^1\Sigma^+_{\rm g}$ and $b$ 
$^3\Sigma^-_{\rm g}$ states, Langhoff et al. (1977) pointed out that such 
interactions are relatively weak, since two electrons are excited.

\subsection{Interstellar C$_2$}

Since the results on C$_2$ based on the $D-X$ band toward the stars 
X Per, 62 Tau, $\rho$ Oph D, $\zeta$ Oph, and HD~207198 have been analyzed 
by us (e.g., Federman et al. 1994; Lambert et al. 1995; Sheffer et al. 2008) 
and others in the past (e.g., van Dishoeck \& Black 1986; Kaczmarczyk 2000; 
Sonnentrucker et al. 2007), here we focus on the new determinations.  We 
consider predictions for the C$_2$ column density inferred from observed 
columns of CH and CN, based on a simple prescription discussed by us earlier 
(e.g., Federman et al. 1994; Knauth et al. 2001).  In particular, we 
compare the predictions in Pan et al. (2005) and Sheffer et al. (2008)  
with the C$_2$ measurements from the present work.  Furthermore, the 
rotational state distribution in terms of column densities yields 
information about the excitation of the molecule, from which gas density 
and temperature and the strength of the infrared radiation field permeating 
the gas can be inferred (van Dishoeck \& Black 1982).

More recent data indicate refinements to 
the analysis of C$_2$ excitation compared to our earlier 
studies.  First, experiments (Erman \& Iwamae 1995) and theoretical 
calculations (Kokkin et al. 2007; Schmidt \& Bacskay 2007) are converging 
on a $f$-value for the $A-X$ (2,0) transition larger than we used in the 
past.  A value of $1.4 \times 10^{-3}$ now seems more appropriate, rather 
than $1.0 \times 10^{-3}$ used by us in the past.
Second, the cross section for excitation 
via collisions is larger than the value suggested by van Dischoek \& Black 
(1982).  More recent quantum mechanical calculations (Lavendy et al. 1991; 
Robbe et al. 1992; Najar et al. 2008, 2009) indicate a cross 
section of $4  \times 10^{-16}$ cm$^2$, rather than $2 \times 10^{-16}$ 
cm$^2$.  

The two changes would partially cancel, yielding $n_{coll}$ 
about 60\% smaller than what we would have obtained with the older values.  
However, a more sophisticated treatment based on the latest information 
and considering core-halo clouds (Casu \& Cecchi-Pestellini 2012) 
suggests that our original method provides a reasonable approximation.  
This arises when a weighted density is obtained for the core and halo 
portions of the clouds seen toward $\zeta$ Oph, which we also studied 
(Lambert et al. 1995).  We, therefore, present results using our earlier 
method with the realization of their approximate nature.  
To extract $n_H$ from $n_{coll}$, we continue to assume there is no 
enhancement in the infrared flux over the average interstellar value and we 
multiply $n_{coll}$ by 1.5.  This accounts for the fact that diffuse 
molecular clouds have roughly equal amounts of atomic and molecular hydrogen.

Table 3 provides comparisons of measured and predicted C$_2$ column 
density, predicted gas densities from CN chemistry and C$_2$ excitation, 
and the gas temperature derived from H$_2$, $T_{01}$(H$_2$), and 
the C$_2$ excitation temperature, $T$(C$_2$), the latter based on 
all observed column densities.  Our column densities appearing in the second 
column are obtained from the detections shown in Table 2; if 
contributions from higher-lying levels were included, $N$(C$_2$) would be 
about 0.1 dex higher.  In the column for gas 
densities from analysis of the CN chemistry, several entries may appear, 
depending on the number of molecular components detected for that sight 
line.  The C$_2$ column densities predicted by the simple CN chemical 
model agree very well with our new determinations.  For HD~206267, 
HD~207308, and HD~208266, the comparison is best when only the main 
molecular component is considered.  This is not surprising; although the 
complete component structure is used to fit the spectra, the other 
components are masked in the noise because the $F-X$ (1,0) band is 
relatively weak.  For HD~206267, we also note that the column density 
inferred from the $F-X$ (0,0), $A-X$ (2,0), and $A-X$ (3,0) bands 
by Sonnentrucker et al. (2007) is in excellent agreement 
with our result.  The gas densities derived from the CN chemistry and 
C$_2$ excitation do not appear to agree very well.  The expected 
agreement is a factor of 2 (e.g., Federman et al. 1994), but the densities 
from excitation are generally lower.  A possible cause for 
the discrepancy may involve the presence of multiple, lower density 
molecular components along the line of sight.  Modeling the spectra may 
favor the dominant component in terms of column density, but the 
distribution of rotational levels may depend on all the components.  
Further evidence for this possibility is found in the comparison between 
$T_{01}$(H$_2$) and $T$(C$_2$).  Except for the gas toward HD~192035, 
which has the highest density from C$_2$ excitation in our sample, the 
two temperatures are comparable.  Earlier studies of molecular rich 
diffuse clouds (see e.g., Sheffer et al. 2007) showed that on average 
$T$(C$_2$) was less than $T_{01}$(H$_2$).  The lower density, higher 
temperature components appear to play a more significant role in the 
determination of C$_2$ excitation.  This idea is supported by the 
more thorough analyses of Casu \& Cecchi-Pestellini (2012).  
We also mention that the excitation 
analysis for the gas toward HD~198781 is not very satisfying.  
Minimizing the differences between the observable distribution and the 
predicted ones could not provide restrictions on $T$(C$_2$).  This 
probably arose because absorption from only $J^{\prime\prime}$ $=$ 4 and 6 
was detected.

\section{Conclusions}

We described a study of interstellar C$_2$ based on archival $HST$ spectra 
of the $D-X$ (0,0), $F-X$ (0,0), and $F-X$ (1,0) bands.  Building on 
earlier work, we provided further details regarding the anomalies in 
line strengths seen in the $F-X$ (0,0) band.  Line shifts were quantified 
for transitions involving $J^{\prime}$ $\le11$.  The lines for both $F-X$ 
bands were found to be broadened by predissociation.  The anomalies likely 
arise from triplet-singlet interactions involving spin-orbit mixing 
between the $F$ $^1\Pi_u$ state and $^3\Pi_{\rm u}$ states combined with 
an avoided crossing between the triplet states.  Other possibilities 
include tunneling through a potential barrier, which is caused by
avoided crossings with the 3 and 4 $^1\Pi_u$ states, and spin-orbit mixing 
between the $F$ state and other triplet states of $ungerade$ symmetry, 
but these are not as likely.  We also determined new C$_2$ column densities 
from the $F-X$ (1,0) band and compared these results with our earlier 
predictions, finding very good agreement.  Our analysis of C$_2$ excitation 
yields somewhat lower gas densities, but these are consistent with core-halo  
models using a refined set of input parameters (Casu \& Cecchi-Pestellini 
2012).  The inferred gas densities are still consistent 
with chemical predictions, when the presence of complex component structure 
along the line of sight having a range of gas densities is considered.

We conclude by noting possible extensions of this work.  First, additional 
theoretical calculations and experiments are needed to elucidate more 
clearly the cause of the anomalies present in $F-X$ spectra.  Second, 
because empirical and theoretical oscillator strengths for the three band 
systems used in interstellar studies ($A-X$, $D-X$, $F-X$) have achieved 
an impressive level of concensus, interstellar C$_2$ abundances are now 
more secure.  Abundances can be determined from any of the bands, but 
analyses should not rely solely on the $F-X$ (0,0) band, where anomalies 
in line strength are the strongest.  Future studies using the $F-X$ 
bands should also incorporate the broadening caused by predissociation 
when modeling the spectra.  This is especially important for studies of 
sight lines where the transition from diffuse molecular to dark clouds is 
taking place.  The abundances are sufficiently large that optical depth 
effects need to be considered.

\acknowledgments
This work was supported by NASA grants NNG 06-GC70G and NNX10AD80G.  
R.C.H. acknowledges support by the National Science Foundation under 
grant 0353899 for the Research Experience for Undergraduates at the
University of Toledo.  We thank Peter Bernath and Pablo Bruna for very 
helpful discussions.  We appreciate the comments of Pablo, Evelyne Roueff, 
and Dan Welty on earlier drafts.  An anonymous referee raised 
insightful points that improved our interpretation of the results 
presented here.  Observations made with the NASA/ESA Hubble Space 
Telescope were obtained from the data archive at STScI.  STScI is 
operated by the Association of Universities for Research in Astronomy, Inc. 
under NASA contract NAS5-26555.

\clearpage

\begin{deluxetable}{ccccccccccccccccc}
\rotate
\tablecolumns{16}
\tablewidth{0pt}
\tabletypesize{\scriptsize}
\tablecaption{Column Densities for C$_2$ $D-X$ (0,0)}
\startdata
\hline\hline
Level & \multicolumn{3}{c}{X Per} & & \multicolumn{2}{c}{62 Tau} & & 
\multicolumn{2}{c}{$\rho$ Oph D} & & \multicolumn{2}{c}{$\zeta$ Oph} & &
\multicolumn{2}{c}{HD 207198} \\ 
\cline{2-4} \cline{6-7} \cline{9-10} \cline{12-13} \cline{15-16}
 & Present & K00\tablenotemark{c} & SWTY07\tablenotemark{d} & & Present & 
SWTY07 & & Present & SWTY07 & & Present & LSF95\tablenotemark{e} & 
 & Present & SWTY07 \\ \hline
$J$ = 0 & 0.165(0.024)\tablenotemark{a}$^,$\tablenotemark{b} & 
0.153(0.051) & 0.15(0.02) & & 0.166(0.012) & 0.15(0.02) & & 0.099(0.010) 
& 0.10(0.02) & & 0.069(0.005) & 0.075(0.015) & & 0.218(0.028) & 0.20(0.02) \\
$J$ = 2 & 0.774(0.051) & 0.648(0.075) & 0.67(0.05) & & 0.594(0.026) 
& 0.57(0.04) & & 0.319(0.021) & 0.33(0.02) & & 0.302(0.012) & 0.290(0.024) & 
 & 0.883(0.175) & 0.83(0.04) \\
$J$ = 4 & 0.784(0.049) & 0.674(0.083) & 0.69(0.04) & & 0.633(0.024) 
& 0.62(0.04) & & 0.367(0.022) & 0.37(0.03) & & 0.356(0.011) & 0.327(0.019) & 
 & 0.984(0.102) & 0.91(0.05) \\
$J$ = 6 & 0.592(0.049) & 0.530(0.069) & 0.55(0.04) & & 0.480(0.023) 
& 0.47(0.03) & & 0.277(0.020) & 0.28(0.02) & & 0.266(0.008) & 0.247(0.019) & 
 & 0.822(0.083) & 0.75(0.04) \\
$J$ = 8 & 0.395(0.046) & 0.346(0.062) & 0.37(0.04) & & 0.331(0.024) 
& 0.32(0.03) & & 0.151(0.023) & 0.18(0.02) & & 0.200(0.010) & 0.192(0.020) & 
 & 0.537(0.072) & 0.51(0.04) \\
$J$ = 10 & 0.293(0.037) & 0.291(0.063) & 0.23(0.03) & & 0.189(0.020) 
& 0.19(0.02) & & 0.128(0.025) & 0.14(0.02) & & 0.145(0.010) & 0.137(0.017) & 
 & 0.335(0.032) & 0.37(0.04) \\
$J$ = 12 & 0.184(0.035) & 0.177(0.056) & 0.22(0.03) & & 0.171(0.021) 
& 0.15(0.02) & & 0.081(0.015) & 0.08(0.02) & & 0.097(0.010) & 0.111(0.014) & 
 & 0.226(0.053) & 0.23(0.04) \\
$J$ = 14 & 0.132(0.037) & 0.139(0.045) & 0.16(0.03) & & 0.106(0.017) 
& 0.11(0.02) & & 0.037(0.015) & 0.05(0.02) & & 0.084(0.009) & 0.077(0.015) & 
 & 0.165(0.057) & 0.17(0.04) \\
$J$ = 16 & 0.094(0.030) & $\ldots$ & 0.04(0.03) & & 0.092(0.019) 
& 0.09(0.02) & & 0.052(0.020) & $\ldots$ & & 0.064(0.008) & 0.062(0.016) & 
 & 0.116(0.024) & 0.13(0.04) \\
$J$ = 18 & 0.060(0.025) & $\ldots$ & $\ldots$ & & 0.067(0.020) 
& 0.06(0.02) & & 0.060(0.020) & $\ldots$ & & 0.049(0.008) & 0.060(0.014) & 
 & 0.109(0.029) & 0.13(0.04) \\
\enddata
\tablenotetext{a}{Column densities in units of 10$^{13}$ cm$^{-2}$.}
\tablenotetext{b}{1 $\sigma$ uncertainties are in parentheses.}
\tablenotetext{c}{Kaczmarczyk 2000.}
\tablenotetext{d}{Sonnentrucker et al. 2007.}
\tablenotetext{e}{\ Lambert et al. 1995.}
\end{deluxetable}

\clearpage

\begin{deluxetable}{cccccccccccc}
\tablewidth{0pt}
\tabletypesize{\tiny}
\tablecaption{Equivalent Widths and Column Densities for C$_2$ $F-X$ (1,0)}
\tablehead{
\colhead{Level} &\colhead{Branch} &\colhead{$W_\lambda$\tablenotemark{a}} &
\colhead{$N_J$\tablenotemark{b}} &\colhead{$W_\lambda$\tablenotemark{a}} &
\colhead{$N_J$\tablenotemark{b}} &\colhead{$W_\lambda$\tablenotemark{a}} &
\colhead{$N_J$\tablenotemark{b}} &\colhead{$W_\lambda$\tablenotemark{a}} &
\colhead{$N_J$\tablenotemark{b}} &\colhead{$W_\lambda$\tablenotemark{a}} &
\colhead{$N_J$\tablenotemark{b}}} 
\startdata
&&\multicolumn{2}{l}{HD 23478} &\multicolumn{2}{l}{HD 147683} &
\multicolumn{2}{l}{HD 177989} &\multicolumn{2}{l}{HD 192035} &
\multicolumn{2}{l}{HD 198781}\\
 & &\multicolumn{2}{l}{$\pm$0.33} &\multicolumn{2}{l}{$\pm$0.38} &
\multicolumn{2}{l}{$\pm$0.35} &\multicolumn{2}{l}{$\pm$0.60} &
\multicolumn{2}{l}{$\pm$0.66} \\
\cline{3-12}
$J$ = 0 &$R$ &1.19 &12.13$\pm$0.11&0.76 &11.94$\pm$0.18&1.78 &
12.35$\pm$0.08&3.00 &12.58$\pm$0.08&\nodata&$<$12.16\\

$J$ = 2 &$R$ &1.58 &12.66$\pm$0.07&1.18 &12.55$\pm$0.10&1.78 &
12.59$\pm$0.09&3.57 &13.06$\pm$0.06&\nodata&$<$12.42\\
        &$Q$ &1.96 &              &1.44 &              &1.28 &              &
4.28 &              &\nodata&   \\
        &$P$ &\nodata&            &\nodata&            &1.56 &              &
\nodata&             &\nodata&   \\

$J$ = 4 &$R$ &1.27 &12.64$\pm$0.07&1.61 &12.78$\pm$0.07&0.95 &
12.53$\pm$0.10&2.13 &12.89$\pm$0.08&1.47 &12.76$\pm$0.12\\
        &$Q$ &1.87 &              &2.27 &              &1.37 &              &
3.05 &              &2.03 &     \\
        &$P$ &\nodata&            &0.86 &              &\nodata&            &
\nodata&             &\nodata&   \\

$J$ = 6 &$R$ &0.67 &12.39$\pm$0.12&1.03 &12.60$\pm$0.09&\nodata&$<$12.20    &
\nodata&12.56$\pm$0.14&1.35 &12.75$\pm$0.12\\
        &$Q$ &1.08 &              &1.59 &              &\nodata&            &
1.56 &              &2.00 &     \\
        &$P$ &\nodata&            &\nodata&            &\nodata&            &
\nodata&             &\nodata&   \\

$J$ = 8 &$R$ &0.74 &12.45$\pm$0.10&1.21 &12.69$\pm$0.08&\nodata&
12.21$\pm$0.18&\nodata&    &\nodata&   \\
        &$Q$ &1.25 &              &1.93 &              &0.70 &              &
\nodata&    &\nodata&   \\
        &$P$ &\nodata&            &\nodata&            &\nodata&            &
\nodata&    &\nodata&   \\
 
$J$ = 10&$R$ &\nodata&            &\nodata&            &\nodata&
12.23$\pm$0.17&\nodata&    &\nodata&   \\
        &$Q$ &\nodata&            &\nodata&            &0.73 &              &
\nodata&    &\nodata&   \\
        &$P$ &\nodata&            &\nodata&            &\nodata&            &
\nodata&    &\nodata&   \\
\cline{3-12}
&&\multicolumn{2}{l}{HD 203532} &\multicolumn{2}{l}{HD 206267} &
\multicolumn{2}{l}{HD 207308} &\multicolumn{2}{l}{HD 208266} &
\multicolumn{2}{l}{HD 220057}\\
 & &$\pm0.55$ & &$\pm0.53$ & &$\pm0.53$ & &$\pm0.42$ & &$\pm0.72$ & \\
\cline{3-12}
$J$ = 0 &$R$ &1.78 &12.35$\pm$0.12&3.50 &12.70$\pm$0.06&2.79 &
12.54$\pm$0.08&2.89 &12.57$\pm$0.06&\nodata&$<$12.14\\

$J$ = 2 &$R$ &2.81 &12.99$\pm$0.07&4.32 &13.23$\pm$0.04&3.55 &
13.06$\pm$0.05&2.07 &12.81$\pm$0.07&\nodata&$<$12.54\\
        &$Q$ &3.31 &              &5.03 &              &4.26 &              &
2.52 &              &\nodata&   \\
        &$P$ &\nodata&            &1.39 &              &\nodata&            &
\nodata&            &\nodata&   \\

$J$ = 4 &$R$ &2.24 &12.95$\pm$0.07&3.76 &13.23$\pm$0.04&3.73 &
13.17$\pm$0.04&1.97 &12.86$\pm$0.06&\nodata&12.50$\pm$0.18\\
        &$Q$ &3.08 &              &4.99 &              &5.14 &              &
2.80 &              &1.38 &     \\
        &$P$ &1.24 &              &2.16 &              &2.05 &              &
1.05 &              &\nodata&   \\

$J$ = 6 &$R$ &1.50 &12.78$\pm$0.09&1.48 &12.77$\pm$0.09&2.08 &
12.91$\pm$0.07&1.36 &12.72$\pm$0.08&\nodata&12.61$\pm$0.15\\
        &$Q$ &2.26 &              &2.26 &              &3.19 &              &
2.11 &              &1.73 &     \\
        &$P$ &\nodata&            &\nodata&            &1.35 &              &
0.88 &              &\nodata&   \\

$J$ = 8 &$R$ &\nodata&12.59$\pm$0.13&\nodata&12.57$\pm$0.13&
1.27 &12.70$\pm$0.10&\nodata&$<$12.27&\nodata&12.66$\pm$0.14\\
        &$Q$ &1.58 &              &1.51 &              &2.07 &              &
\nodata&            &1.92 &     \\
        &$P$ &\nodata&            &\nodata&            &\nodata&            &
\nodata&            &\nodata&   \\

$J$ = 10&$R$ &\nodata&12.65$\pm$0.12&\nodata&            &1.15 &
12.67$\pm$0.10&\nodata&12.42$\pm$0.14&\nodata&$<$12.43     \\
        &$Q$ &1.75 &              &\nodata&            &1.94 &              &
1.13 &              &\nodata&   \\
        &$P$ &\nodata&            &\nodata&            &\nodata&            &
\nodata&            &\nodata&   \\

$J$ = 12&$R$ &\nodata&12.47$\pm$0.16&\nodata&            &\nodata&
12.61$\pm$0.12&\nodata&   &\nodata&$<$12.44     \\
        &$Q$ &1.22 &              &\nodata&            &1.72 &              &
\nodata&   &\nodata&   \\
        &$P$ &\nodata&            &\nodata&            &\nodata&            &
\nodata&   &\nodata&   \\
\enddata
\tablenotetext{a}{Equivalent width in m\AA. A constant 1-$\sigma$ uncertainty 
in $W_\lambda$ is given for each sight line.}
\tablenotetext{b}{\ Log column density in cm$^{-2}$. Upper limits on $N_J$ 
are 2-$\sigma$.}
\end{deluxetable}

\clearpage

\begin{deluxetable}{lcccccc}
\tablecaption{Comparison of Results}
\tablecolumns{7}
\tablewidth{0pt}
\tabletypesize{\scriptsize}
\tablehead{
\colhead{Star} & \colhead{log $N$(C$_2$)\tablenotemark{a}} & 
\colhead{log $N_p$(C$_2$)\tablenotemark{b}} & 
\colhead{$n_H$(CN)\tablenotemark{b}} & \colhead{$n_H$(C$_2$)} & 
\colhead{$T_{01}$(H$_2$)} & \colhead{$T$(C$_2$)}\\
 & & & (cm$^{-3}$) & (cm$^{-3}$) & (K) & (K)}
\startdata
HD~23478 & 13.19(0.09) & 13.18\tablenotemark{c} & 325, 775\tablenotemark{c} & 
190 & 55\tablenotemark{c} & 20--40 \\
HD~147683 & 13.29(0.11) & $\ldots$ & $\ldots$ & 95 & 58\tablenotemark{c} & 
50--60 \\
HD~177989 & 13.11(0.13) & $\ldots$ & $\ldots$ & 210 & 49\tablenotemark{c} & 
20 \\
HD~192035 & 13.43(0.10) & 13.52\tablenotemark{c} & 
$\le$575, 1550 $\le$725\tablenotemark{c} & 
350 & 68\tablenotemark{c} & 20--30 \\
HD~198781 & 13.06(0.16) & 13.12\tablenotemark{c} & 750\tablenotemark{c} & 
$\le$95 & 65\tablenotemark{c} & 10--100 \\
HD~203532 & 13.58(0.09) & $\ldots$ & $\ldots$ & 210 & 47\tablenotemark{c} & 
20--40 \\
HD~206267 & 13.69(0.06) & 13.81\tablenotemark{d} & 
80, 1150, 525, 60 \tablenotemark{d} & 280 & 65\tablenotemark{d} & 
20--40 \\
 & 13.68(0.04)\tablenotemark{e} & 13.70\tablenotemark{f} & \\
HD~207308 & 13.71(0.07) & 13.90\tablenotemark{d} & 
600, 300\tablenotemark{d} & 190 & 57\tablenotemark{d} & 
20--50 \\
 & $\ldots$ & 13.79\tablenotemark{f} & \\
HD~208266 & 13.40(0.09) & 13.66\tablenotemark{d} & 
375, 90, 125\tablenotemark{d} & 210 & $\ldots$ & 10 \\
 & $\ldots$ & 13.52\tablenotemark{f} & \\
HD~220057 & 13.07(0.22) & 13.07\tablenotemark{c} & 
750, $\le$850\tablenotemark{c} & 165 & 65\tablenotemark{c} & 
90 \\
\enddata
\tablenotetext{a}{Numbers in parentheses are uncertainties inferred from 
those for the two strongest lines, taken in quadrature.}
\tablenotetext{b}{Predictions based on our chemical model for CN and 
reproduced from Pan et al. 2005 and Sheffer et al. 2008.}
\tablenotetext{c}{From Sheffer et al. 2008.}
\tablenotetext{d}{From Pan et al. 2005.}
\tablenotetext{e}{Sonnentrucker et al. 2007 from fitting the $F-X$ (0,0), 
$A-X$ (2,0), and $A-X$ (3,0) bands.}
\tablenotetext{f}{Main component given in Pan et al. 2005.}
\end{deluxetable}

\clearpage

\begin{figure}
\hspace{0.3in}
\includegraphics[scale=0.80]{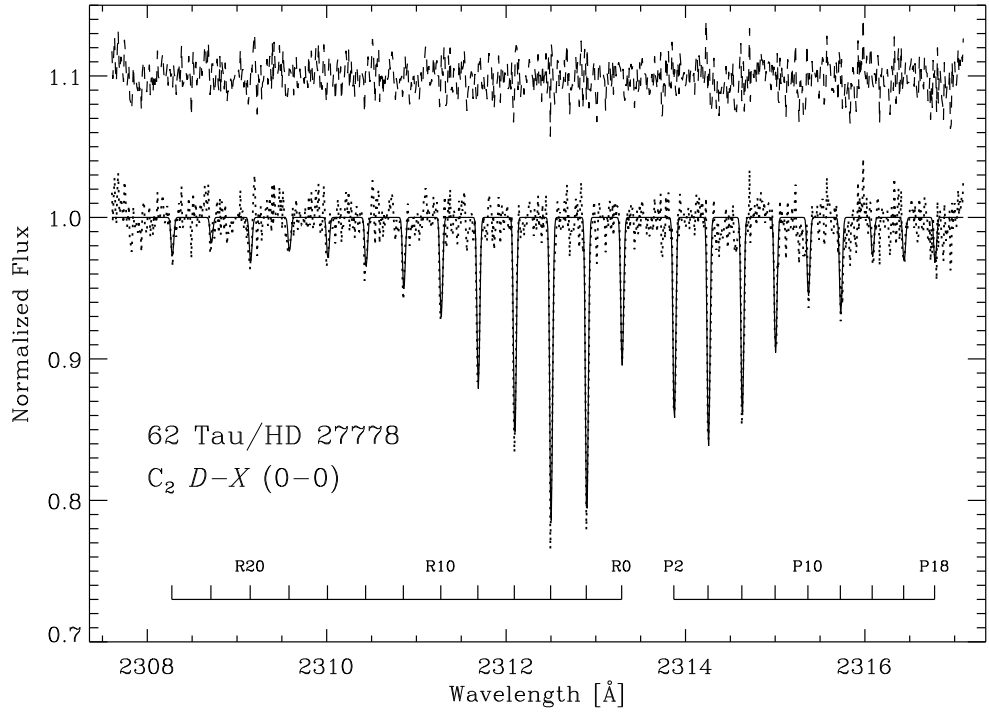}
\vspace{0.3in}
\caption{Absorption from the $D-X$ band of C$_2$ toward HD~27778.  The 
dotted line indicates data normalized to unity, while the solid line is the 
fit based on component structure used by Sheffer et al. (2008) for CO.  
Individual transitions are identified at the bottom of the figure.  
The residuals appear above the spectrum at 1.1 as a dashed line.}
\end{figure}

\clearpage

\begin{figure}
\hspace{0.2in}
\includegraphics[scale=1.00]{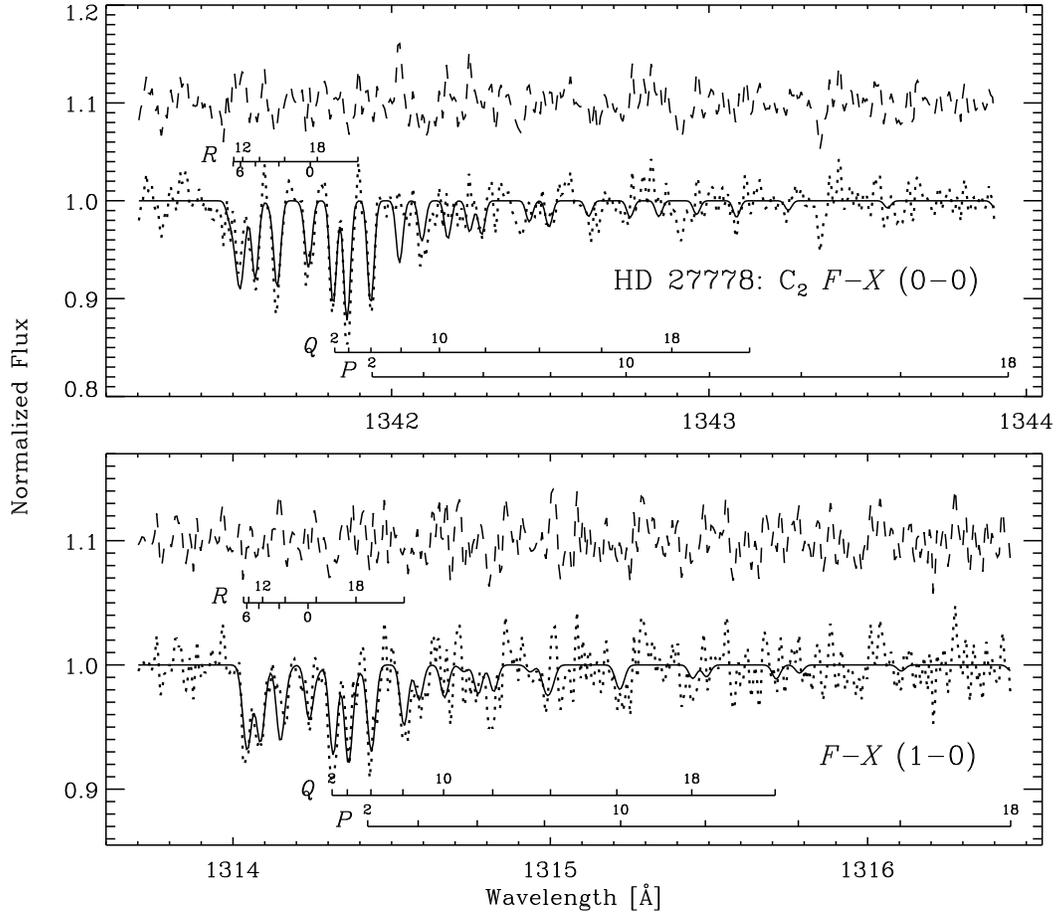}
\vspace{0.3in}
\caption{As Fig. 1 for absorption from the $F-X$ bands of C$_2$ toward 
HD~27778.  The upper panel shows the fit to the (0,0) band and the lower 
panel that for the (1,0) band.  The fits were based on the column density 
derived from the $D-X$ band as presented in Fig. 1.}
\end{figure}

\clearpage

\begin{figure}
\hspace{0.7in}
\includegraphics[scale=1.20]{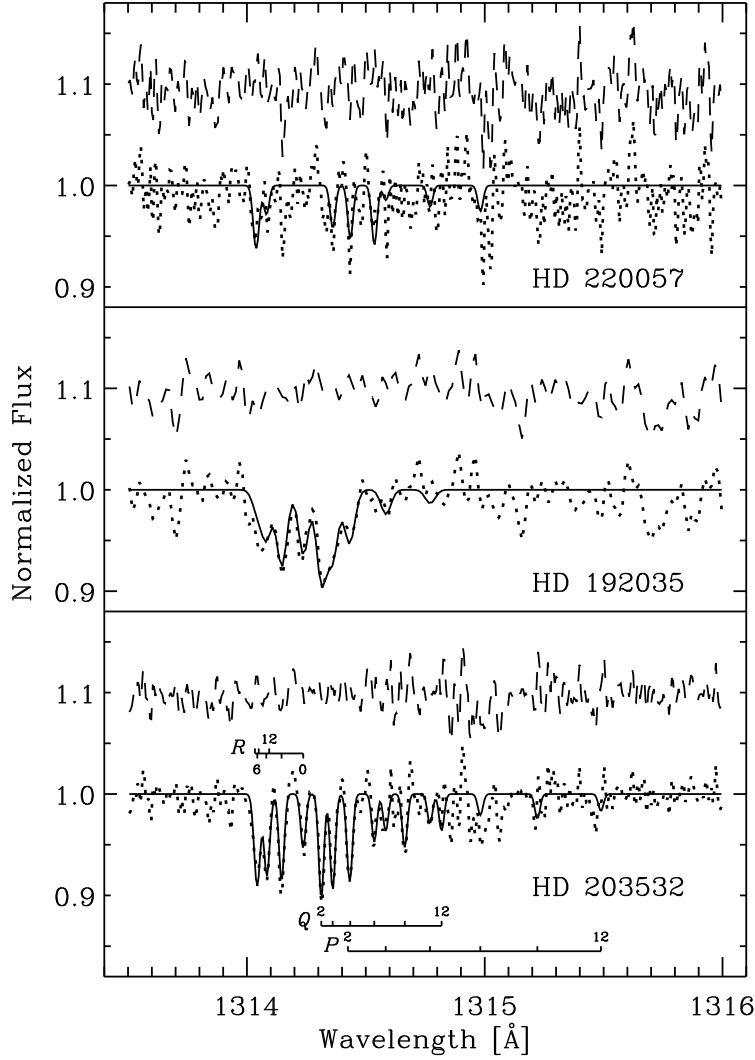}
\vspace{0.3in}
\caption{As Fig. 1 for absorption from the $F-X$ (1,0) band of C$_2$ toward
HD~220057, HD~192035, and HD~203532, a sample of stars revealing new 
detections of C$_2$.  The three stars are arranged from top to bottom 
according to $N$(C$_2$).  Models show absorption from $J$ levels that are 
securely determined as presented in Table 2.  
Spectra of HD~220057 and HD~192035 have been 
shifted by $-0.120$ and $-0.128$ \AA, respectively, to match the wavelength 
scale of HD~203532.  HD~192035 was observed with the lower-resolution 
grating E140M.}
\end{figure}

\end{document}